\documentclass{article}

\usepackage[preprint]{neurips_2024}
\setcitestyle{numbers,square,comma}

\usepackage[utf8]{inputenc}
\usepackage[T1]{fontenc}
\usepackage{amsmath,amssymb,amsfonts}
\usepackage{bm}
\usepackage{graphicx}
\usepackage{booktabs}
\usepackage{multirow}
\usepackage{hyperref}
\hypersetup{colorlinks=true,linkcolor=blue,citecolor=blue,urlcolor=blue}
\usepackage{cleveref}
\usepackage{float}
\usepackage{algorithm}
\usepackage{algpseudocode}
\usepackage{tikz}
\usetikzlibrary{arrows.meta,positioning,calc,shapes.geometric,fit,backgrounds}
\usepackage{pgfplots}
\pgfplotsset{compat=1.18}
\usepackage{enumitem}
\usepackage{subcaption}
\usepackage{pifont}
\newcommand{\cmark}{\textcolor{green!60!black}{\ding{51}}}
\newcommand{\xmark}{\textcolor{red}{\ding{55}}}

\title{Binding Drift in Multi-Step Tool-Augmented Agents}

\author{
  Rahul Suresh Babu\thanks{Equal contribution.}\\
  \texttt{rahulsb@bu.edu}
  \And
  Shashank Indukuri\footnotemark[1]\\
  \texttt{sinduku1@depaul.edu}
}

\begin{document}
\maketitle

\begin{abstract}
Tool-augmented language-model agents execute multi-step workflows over external systems, resolving an entity once and then acting on it across subsequent steps. Prior work shows that in single-step actions, agents select the correct tool but bind it to the wrong entity 24--26\% of the time~\cite{babu2026entitybinding}. We study what happens to entity bindings \emph{over time}: do they stay correct, silently drift to a different entity, or, if wrong from the start, propagate and compound? We formalize \emph{binding drift} (correct at step~1, wrong later) as distinct from \emph{error propagation} (wrong at step~1, carried forward), and score them on disjoint workflow sets so the two cannot be conflated. In a controlled multi-step testbed (200 workflows, 580 entity-binding-scored steps, four enterprise domains, eight model backends spanning small to frontier), we find: (1)~under controlled error injection, an entity lock (the intuitive ``persist the first binding'' fix) amplifies wrong actions from 907 to 2{,}746 ($3.0\times$; bootstrap 95\% CI $[2.8, 3.3]$), because it faithfully carries the seeded wrong entity into every later step; (2)~the amplification reaches $8.5\times$ on the most affected model (Claude Opus~4.5); (3)~a practical LLM-based re-verifier (a single cheap second model call re-reading the original instruction) reduces wrong actions by 79\% ($0.21\times$; CI $[0.18, 0.25]$), closing the gap to within 1 percentage point of an oracle upper-bound ($0.20\times$); and (4)~in the natural (non-injected) setting, baseline agents drift on 18\% of eligible workflows, with the per-step error rate rising across steps. Persistence and re-verification are not interchangeable: a defense that eliminates drift can worsen propagation, and a practical re-verifier nearly matches oracle recovery.
\end{abstract}

\section{Introduction}
\label{sec:intro}

Tool-augmented language-model agents rarely act once. A realistic task is a \emph{workflow}: pull up an account, log a note on it, update its renewal date, then email its owner a summary. Every step after the first refers back to the same real-world entity, often only as ``it,'' ``that account,'' or ``the ticket.'' The agent must not only resolve the entity correctly the first time, but keep acting on the \emph{same} entity for the rest of the workflow.

Prior work established that agents choose the right tool but bind it to the wrong entity in a single action~\cite{babu2026entitybinding}. That work measured a point failure: one instruction, one action, scored once. It could not observe what happens across steps. This paper studies the trajectory.

We identify two multi-step failure modes invisible to single-step evaluation:
\begin{itemize}[nosep]
\item \textbf{Binding drift.} The agent binds the correct entity at step~1, then a later step silently acts on a \emph{different} valid entity because a pronoun is re-resolved, a same-named distractor becomes salient, or the early resolution scrolls out of context.
\item \textbf{Error propagation.} The agent binds the \emph{wrong} entity at step~1 and carries it forward, so one mis-binding becomes a wrong action at every subsequent step.
\end{itemize}

These are different problems with different fixes. A persistence mechanism (entity lock) eliminates drift but, as we show, amplifies propagation when the initial binding is wrong. An independent re-derivation mechanism can in principle address both (as our oracle upper-bound demonstrates), but at an over-clarification cost.

\paragraph{Relationship to dialogue state tracking.} The concept of persisting a slot value across turns is well-studied in dialogue state tracking (DST) systems such as MultiWOZ~\cite{budzianowski2018multiwoz} and SGD~\cite{rastogi2020sgd}. In those settings, slot persistence is assumed to be correct behavior. Our finding inverts this assumption: in agentic tool-use, where each resolution triggers an irreversible real-world action, slot persistence is actively harmful when the initial fill is wrong. This inversion (from ``persistence is good'' to ``persistence can amplify errors'') is one specific contribution of this work to the broader state-tracking literature.

\paragraph{Contributions.}
\begin{itemize}[nosep]
\item A formalization that scores drift and propagation on disjoint workflow sets, making them independently measurable (\Cref{sec:formal}).
\item A taxonomy of five drift-inducing patterns and a controlled multi-step testbed (\Cref{sec:bench}).
\item A controlled-injection protocol that decouples propagation measurement from natural error rate by seeding known-wrong bindings, enabling comparison across models with different baseline failure rates (\Cref{sec:results}).
\item An empirical finding that entity locking amplifies propagation up to $8.5\times$, and a practical LLM-based re-verifier (a single cheap second call) that achieves 79\% reduction, closing to within 1 percentage point of the oracle ceiling (\Cref{sec:results}).
\end{itemize}

\section{Problem Formulation}
\label{sec:formal}

\paragraph{Workflows and binding state.}
A workflow is an ordered sequence of steps $w = (s_1, \ldots, s_K)$ executed over an environment store $\mathcal{E}$ of typed entities (persons, accounts, events, documents, tickets). At each step $s_k$ the agent emits an action $a_k = \langle t_k, \hat{b}_k \rangle$: a tool $t_k$ and a predicted entity binding $\hat{b}_k \in \mathcal{E} \cup \{\bot\}$.

Rather than assuming a single gold entity per workflow, we define a \emph{binding-state map} that captures the full ground truth at each step:
\[
B_k = \{\sigma_1 \mapsto e_1^{(k)},\; \sigma_2 \mapsto e_2^{(k)},\; \ldots,\; \sigma_m \mapsto e_m^{(k)}\}
\]
where each $\sigma_i$ is a referent \emph{slot} (a semantic role the workflow refers to, such as ``the account'' or ``the account's owner'') and $e_i^{(k)} \in \mathcal{E}$ is the gold entity filling that slot at step $k$. Every scored action step has exactly one \emph{active slot} $\sigma(k)$ whose gold entity is the correct binding for that step: $e_k^\star = B_k[\sigma(k)]$.

\paragraph{Primary carry slot vs.\ derived slots.}
We distinguish two slot types:
\begin{itemize}[nosep]
\item The \textbf{primary carry slot} $\sigma_1$ is the referent that persists across the majority of steps (e.g., ``the account''). Its gold entity $e_1^{(k)}$ is constant throughout the workflow: $e_1^{(k)} = e_1^\star$ for all $k$.
\item \textbf{Derived slots} $\sigma_2, \ldots, \sigma_m$ hold entities that are determined by the primary slot through a known relation (e.g., $\sigma_2 = \texttt{owner-of}(\sigma_1)$, so $e_2^\star = \texttt{owner-of}(e_1^\star)$). Derived slots become active only at specific steps.
\end{itemize}
Most workflows in our benchmark are single-slot ($m{=}1$): every step acts on $e_1^\star$. The dependent-chain pattern (D3) uses a derived slot: steps 1--3 act on the primary (an account), and step 4 acts on a derived entity (that account's owner, a different entity type). The binding-state map makes this explicit without requiring a separate formalism for cross-domain steps.

\paragraph{Scoring condition.}
Following~\cite{babu2026entitybinding}, entity correctness is scored only when the agent selected the correct tool, isolating binding errors from tool-selection errors. Let
\[
A(w) = \{k : d_k = \mathrm{ACT} \;\wedge\; t_k = t_k^\star\}
\]
be the set of \emph{action steps} where the agent acted with the correct tool. Steps where the agent clarified ($d_k = \mathrm{CLARIFY}$), deferred, or produced malformed output are excluded from the binding-error numerator but tracked separately.

\paragraph{Drift, propagation, and recovery.}
Let $k_1 = \min A(w)$ be the first scored action step. All three indicators are conditioned on the primary carry slot $\sigma_1$, since it is the slot that persists and whose early-binding outcome determines whether errors compound:
\begin{align}
\mathrm{Drift}(w) &= \mathbf{1}\!\left[\hat{b}_{k_1} = e_1^\star \;\wedge\; \exists\, k \in A(w),\, k > k_1 : \hat{b}_k \neq e_k^\star \right] \label{eq:drift}\\
\mathrm{Propagation}(w) &= \mathbf{1}\!\left[\hat{b}_{k_1} \neq e_1^\star \;\wedge\; \textstyle\sum_{k \in A(w)} \mathbf{1}[\hat{b}_k \neq e_k^\star] > 1 \right] \label{eq:prop}\\
\mathrm{Recovery}(w) &= \mathbf{1}\!\left[\hat{b}_{k_1} \neq e_1^\star \;\wedge\; \exists\, k \in A(w),\, k > k_1 : \hat{b}_k = e_k^\star \right] \label{eq:recov}
\end{align}
Drift and propagation are scored on \emph{disjoint} workflow sets: drift applies only to workflows that were correctly bound at $k_1$; propagation applies only to workflows that were incorrectly bound at $k_1$. An agent cannot improve one metric at the cost of the other. This separation is what makes the lock-amplification finding (\Cref{sec:results}) unambiguous: the lock eliminates drift on one set while worsening propagation on a different set.

\paragraph{Derived-binding corruption.}
When the primary carry slot is wrong, derived slots inherit the error through their provenance relation. Concretely, if the agent mis-resolves the account to $\tilde{e}_1 \neq e_1^\star$ at step~1, then at step~4 the derived lookup $\texttt{owner-of}(\tilde{e}_1)$ returns the wrong person. This is the cross-domain cascade visible in D3 workflows: one wrong primary binding corrupts a downstream entity of a different type, and the agent has no local signal that anything is wrong because the derived resolution is internally consistent given the (incorrect) carried state. We call this \emph{derived-binding corruption}; it is the mechanism that produces D3's extreme lock-amplification ratio.

\paragraph{Metrics.} Beyond the three workflow-level indicators, we report:
\begin{itemize}[nosep]
\item \textbf{Per-step wrong-entity rate.} $|\{k \in A(w) : \hat{b}_k \neq e_k^\star\}| \;/\; |A(w)|$ at each step index, averaged across workflows. Reveals the drift curve across the trajectory.
\item \textbf{Compounding factor.} Mean wrong-entity actions per affected workflow (workflows with at least one wrong binding). Quantifies cascade severity.
\item \textbf{Over-clarification rate.} Fraction of resolvable steps where the agent clarified unnecessarily. Captures the safety/completion tension introduced by re-verification defenses.
\item \textbf{Workflow success.} All steps in $A(w)$ have $\hat{b}_k = e_k^\star$. The strictest aggregate: one wrong binding anywhere fails the workflow.
\end{itemize}

\section{Benchmark}
\label{sec:bench}

\subsection{Drift Patterns (D1--D5)}

We construct workflows around five stress patterns derived from common enterprise failure scenarios. Table~\ref{tab:patterns} summarizes the design; Appendix~\ref{app:examples} shows representative workflows verbatim.

\begin{table}[t]
\caption{Drift-pattern taxonomy. Each pattern stresses a different mechanism by which a later step might bind the wrong entity. $N_w$ = workflows per pattern.}
\label{tab:patterns}
\centering
\begin{tabular}{llrl}
\toprule
ID & Pattern & $N_w$ & Pressure mechanism \\
\midrule
D1 & Temporal recurrence & 40 & Multiple instances of a recurring event; name suppressed after step~1 \\
D2 & Distractor injection & 40 & Same-named entity with stronger metadata injected mid-workflow \\
D3 & Dependent chain & 40 & Step $k$ references entity resolved at step $k{-}1$ (cross-domain) \\
D4 & True ambiguity & 40 & No disambiguator; correct behavior is to clarify \\
D5 & Silent switch & 40 & Three same-named entities; qualifier suppressed after step~1 \\
\midrule
& \textbf{Total} & \textbf{200} & \textbf{890 total steps (580 entity-binding-scored)} \\
\bottomrule
\end{tabular}
\end{table}

\subsection{Environment}

The testbed uses four enterprise domains (calendar, CRM, docs, tickets) with cross-domain email actions in D3 workflows with the same entity schema as the single-step benchmark~\cite{babu2026entitybinding}: entities have a unique ID, type (person / document / event / account / ticket), name, and metadata. Each workflow has a primary carry entity across 3--5 steps (with D3 workflows additionally referencing a derived entity at their final step; see \Cref{sec:formal}). Workflows include both \emph{resolve steps} (search/lookup with no binding scored, used to establish context and carry the referent forward) and \emph{action steps} (state-changing tool calls with a gold entity binding, which are the only steps that contribute to drift/propagation metrics). All data is synthetic; no real or proprietary data is used. The generator, environment, harness, and per-step trajectory logs are released.

\subsection{Controlled Error Injection}
\label{sec:injection}

Because capable models rarely mis-bind on our tasks unaided (consistent with~\cite{babu2026entitybinding}'s finding that wrong-entity errors concentrate in temporal and true-ambiguity conditions; our non-ambiguity workflows produce lower unaided rates), natural-mode denominators for propagation are small. To measure propagation behavior at scale, we \emph{seed} a wrong early binding into the workflow state and history, then observe whether each defense lets the error propagate or corrects it. Concretely:
\begin{enumerate}[nosep]
\item A designated distractor entity ID (annotated per workflow) is injected as the ``resolved'' binding in the workflow state before step~1.
\item The model's conversation history includes a synthetic prior turn: ``(prior) identified the [slot] as [wrong\_id].''
\item If the entity-lock feature is active, the lock is pre-set to the wrong ID.
\end{enumerate}
This gives a large, capability-independent sample and isolates propagation from initial failure rate, analogous to fault injection in systems testing~\cite{natella2016}. Every workflow (excluding true-ambiguity controls) has an annotated injection target.

\section{Defenses}
\label{sec:defense}

Each defense is deterministic structure outside the model, operating on workflow state:
\begin{itemize}[nosep]
\item \textbf{Entity lock.} Pin the first concrete binding; later steps reuse it. The intuitive ``resolve once, carry forward'' fix.
\item \textbf{Provenance checkpoint.} Record (referent $\to$ bound ID $\to$ evidence) at each step.
\item \textbf{Drift-aware gate.} If a later step's resolved entity diverges from the lock, clarify instead of acting.
\item \textbf{Independent re-verification (oracle).} Re-derive the target from the original request using a ground-truth resolver; if the carried binding disagrees, correct it. This is an \emph{upper bound}: it shows the ceiling of what a deployable re-verifier could achieve, not a shipping system.
\end{itemize}

The critical contrast: \emph{persistence} (trust the first binding) vs.\ \emph{re-derivation} (recompute from the original request). Persistence is safe only if the first binding was correct.

\begin{algorithm}[t]
\caption{Entity-lock with drift-aware gate for step $s_k$}
\label{alg:lock}
\begin{algorithmic}[1]
\Require locked ID $\ell$ (or $\bot$), raw model binding $\hat{b}_k$, decision $d_k$
\If{$d_k = \textsc{Act}$ and $\ell \neq \bot$ and $\hat{b}_k \neq \ell$}
  \State $d_k \gets \textsc{Clarify}$ \Comment{divergence from lock: re-confirm, don't switch}
\ElsIf{$d_k = \textsc{Act}$ and $\ell = \bot$ and $\hat{b}_k \neq \bot$}
  \State $\ell \gets \hat{b}_k$ \Comment{set lock on first concrete binding}
\EndIf
\State \textbf{provenance:} append $(k, \hat{b}_k, d_k, \text{evidence})$
\If{reverify enabled and $d_k = \textsc{Act}$}
  \State $e_{\mathrm{fresh}} \gets \textsc{Resolve}(\text{original referent}, \mathcal{E})$ \Comment{oracle resolution}
  \If{$\hat{b}_k \neq e_{\mathrm{fresh}}$}
    \State $\hat{b}_k \gets e_{\mathrm{fresh}}$; $\ell \gets e_{\mathrm{fresh}}$ \Comment{correct the carried binding}
  \EndIf
\EndIf
\end{algorithmic}
\end{algorithm}

\section{Evaluation}
\label{sec:results}

\subsection{Setup}

Eight model backends spanning small to frontier: Amazon Nova Micro, Llama-3.1-8B Instruct, Amazon Nova~2 Lite, Claude Haiku~4.5, Llama-3.3-70B Instruct, Amazon Nova Premier, Claude Sonnet~4.5, and Claude Opus~4.5. Five defense configurations for the main injection experiment (baseline, entity lock, LLM self-reverify, LLM cross-reverify, oracle re-verify) plus a persistence stack (lock + provenance + drift gate) evaluated in natural mode. Two experimental conditions: \emph{natural} (agent resolves each step itself) and \emph{controlled injection} (wrong early binding seeded per \Cref{sec:injection}). All models accessed via Amazon Bedrock Converse API (\texttt{us-east-1}), provider-default decoding (temperature not explicitly set; greedy or near-greedy depending on model), \texttt{maxTokens}=700, sequential calls with 0.15--0.8s inter-request delay. Per-step trajectory logs for all scored runs are released (approximately 40{,}000 model calls across injection and natural conditions, plus retries).

\paragraph{Scoring of malformed outputs.} Steps where the model produced unparseable JSON (PARSE\_ERROR) or where the API returned a transient error are excluded from entity-correctness scoring but counted as workflow-level failures. Across all runs, 1.2\% of steps were errors or parse failures (concentrated in the smallest models). Steps with wrong tool selection are excluded from the entity-binding numerator (following~\cite{babu2026entitybinding}) but count as failures for workflow success.

\subsection{Injection Results}

Table~\ref{tab:inject} reports the controlled-injection experiment (1{,}280 eligible workflows per config = 8 models $\times$ 160 non-ambiguity workflows). The entity lock amplifies wrong actions from 907 (baseline) to 2{,}746 ($3.0\times$; bootstrap 95\% CI $[2.8, 3.3]$)\footnote{Bootstrap CIs computed via 10{,}000 resamples of (workflow, model) pairs with replacement, computing the config-ratio on each resample. Percentile method, seed=42.} by faithfully pinning the seeded wrong entity. A practical LLM cross-verifier (a single cheap second call using Nova Micro) reduces them to 189 ($0.21\times$; CI $[0.18, 0.25]$), comparable to the oracle's 180 ($0.20\times$; CI $[0.17, 0.23]$).\footnote{The cross-verifier outperforms the oracle on some models because the oracle corrects only to the primary carry-slot entity, while the LLM re-verifier can independently resolve derived entities (e.g., ``that account's owner'') that the rigid oracle does not handle. The oracle is thus an approximate upper bound, not a strict one, for workflows with derived slots.}

\begin{table}[t]
\caption{Controlled-injection experiment (200-workflow benchmark): downstream wrong-entity actions after a seeded early mis-binding, by defense. N=1{,}280 eligible workflows (8 models $\times$ 160 non-ambiguity) per row. The entity lock amplifies ($3.0\times$); a practical LLM cross-verifier nearly matches the oracle.}
\label{tab:inject}
\centering
\begin{tabular}{lrrrr}
\toprule
Defense & Wrong actions & $\times$ baseline & Over-clar. & WF succ.\,\% \\
\midrule
Baseline (no defense) & 907 & 1.00 & 743 & 29.2 \\
Entity lock & 2{,}746 & 3.03 & 1{,}040 & 0.0 \\
LLM self-reverify (same model) & 261 & 0.29 & 2{,}044 & 26.5 \\
LLM cross-reverify (Nova Micro) & 189 & 0.21 & 2{,}056 & 25.2 \\
Oracle re-verify (upper bound) & 180 & 0.20 & 2{,}400 & 18.8 \\
\bottomrule
\end{tabular}
\end{table}

\subsection{Per-Model Heterogeneity}

Table~\ref{tab:ladder} breaks the injection result down by model. The lock's amplification ratio varies across models, with the highest values on Claude Opus~4.5 ($8.5\times$), Nova Premier ($5.3\times$), Llama-3.3-70B ($3.3\times$), and Claude Sonnet~4.5 ($2.6\times$). These models propagate the locked binding at high rates under injection. One plausible explanation is that they treat carried workflow state as authoritative; if so, this property becomes a liability when the carried state is wrong. On Claude Haiku~4.5 the lock amplifies modestly ($3.9\times$). The cross-verifier (Nova Micro as cheap second call) reduces wrong actions to near-oracle levels on every model, confirming that practical re-verification is model-independent.

\begin{table}[t]
\caption{Wrong-entity actions per model under injection (N=160 eligible workflows per cell). The lock amplifies on all 8 models. The cross-verifier (cheap second call) achieves near-oracle recovery across the board.}
\label{tab:ladder}
\centering
\begin{tabular}{lrrrrl}
\toprule
Model & Baseline & Lock & Cross-verify & Oracle & Lock effect \\
\midrule
Nova Micro & 208 & 376 & 1 & 7 & $1.8\times$ \\
Llama-3.1-8B & 73 & 190 & 27 & 0 & $2.6\times$ \\
Nova~2 Lite & 191 & 498 & 38 & 40 & $2.6\times$ \\
Claude Haiku~4.5 & 14 & 54 & 2 & 20 & $3.9\times$ \\
Llama-3.3-70B & 135 & 452 & 39 & 24 & $3.3\times$ \\
Nova Premier & 46 & 242 & 35 & 35 & $5.3\times$ \\
Claude Sonnet~4.5 & 188 & 492 & 30 & 33 & $2.6\times$ \\
Claude Opus~4.5 & 52 & 442 & 17 & 21 & $8.5\times$ \\
\bottomrule
\end{tabular}
\end{table}

\subsection{Natural-Mode Drift}

\begin{figure}[t]
\centering
\begin{tikzpicture}
\begin{axis}[width=10cm,height=6.5cm,xlabel={Action step index},
  ylabel={Wrong-entity action rate (\%)},legend pos=north east,
  xtick=data,ymin=0,grid=both]
\addplot coordinates {(1,8.0) (2,7.6) (3,28.1)};
\addlegendentry{Baseline}
\addplot coordinates {(1,8.7) (2,8.3) (3,5.5)};
\addlegendentry{+Lock}
\addplot coordinates {(1,6.6) (2,6.6) (3,6.2)};
\addlegendentry{+Drift gate}
\addplot coordinates {(1,5.6) (2,5.6) (3,5.5)};
\addlegendentry{Persistence stack}
\end{axis}
\end{tikzpicture}
\caption{Per-step wrong-entity rate (natural mode, eight models aggregated). Baseline rises sharply at step~3; the lock flattens the curve. N=288 at steps 1--2; N=128 at step~3.}
\label{fig:curve}
\end{figure}
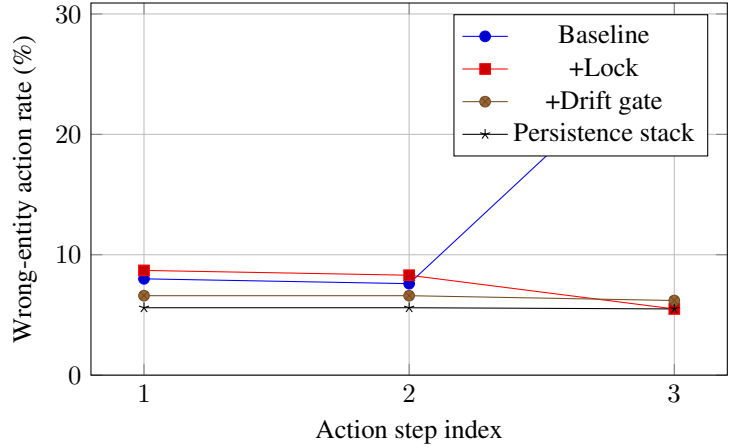

Without injection, baseline agents drift on 18\% of eligible workflows (those correctly bound at step~1, 8 models aggregated). Figure~\ref{fig:curve} shows the per-step wrong-entity rate rising from ${\sim}8\%$ at steps~1--2 to 28\% at step~3. Because not all workflows have a third action step, step~3 uses a smaller subset (N=128 vs.\ N=288 at steps~1--2). On the fixed cohort of 128 workflows with three or more action steps, the pattern holds: 3\% at steps~1--2, rising to 28\% at step~3. The entity lock eliminates this rise, holding the rate flat below 9\%.

\subsection{The Two Failure Modes Respond Oppositely}

The lock eliminates drift (0\%) but, under controlled injection, amplifies propagation ($3.0\times$). This illustrates the formalization's value: scoring on disjoint sets reveals that a defense optimized for one failure mode can worsen the other. The practical implication is that practitioners cannot deploy a single persistence mechanism and assume both problems are solved; re-derivation (as bounded by our oracle experiment) addresses propagation where persistence cannot.

\subsection{Over-Clarification and the Safety--Completion Tradeoff}

Every defense trades wrong-action reduction against over-clarification (743 steps at baseline to 2{,}400 under oracle re-verify in the injection condition). Workflow success ranges from 0\% (lock, which pins the wrong entity and never succeeds) to 29.2\% (baseline), 25.2\% (cross-verifier), and 18.8\% (oracle, which over-clarifies more due to forced re-derivation). On these deliberately hard tasks, the safe response to unresolved ambiguity is to ask rather than act. This mirrors the safety--completion tradeoff of prior single-step work~\cite{babu2026entitybinding}.

\section{Related Work}
\label{sec:related}

\paragraph{Single-step entity binding and tool-use reliability.}
\citet{babu2026entitybinding} isolated the correct-tool/wrong-entity failure mode in single actions, showing 24--26\% wrong-entity rates even when tool selection is correct. We extend along the trajectory dimension, formalizing drift and propagation as distinct failure modes with different causes and fixes. Our controlled-injection protocol further decouples propagation measurement from natural error rate. Other tool-use reliability work~\cite{babu2026toolchoiceconfusion,patil2023gorilla,qin2023toolllm,schick2023toolformer} focuses on tool selection; our contribution is orthogonal: we hold tool selection constant and study entity identity across steps.

\paragraph{Dialogue state tracking and slot persistence.}
Slot persistence is the central mechanism of dialogue state tracking (DST). MultiWOZ~\cite{budzianowski2018multiwoz} and SGD~\cite{rastogi2020sgd} model belief states that accumulate across turns. Neural DST architectures such as TRADE~\cite{wu2019trade} (copy-mechanism generation) and TripPy~\cite{heck2020trippy} (triple copy from span, system-inform, and carry-over) all treat persistence as correct behavior. Our finding inverts this for the agentic setting: when each resolution triggers an irreversible action, persistence amplifies wrong actions by $3.0\times$ under injection. The entity lock is formally equivalent to DST carry-over; we show it cascades errors that DST systems can recover from (via re-prompting) but agentic systems cannot (actions are irreversible).

\paragraph{Entity linking and coreference resolution.}
Entity linking maps mentions to knowledge-base entries. BLINK~\cite{wu2020blink} introduced dense bi-encoder retrieval; GENRE~\cite{decao2021genre} recast it as autoregressive name generation; \citet{sevgili2022neuralel} surveyed the neural EL landscape. Coreference resolution~\cite{lee2017coref} determines which mentions refer to the same entity within static text. Our setting differs: resolution is \emph{incremental} (the agent resolves at step~1, then must maintain that resolution across future steps), the entity store may change between steps, and each resolution is \emph{consequential} (a wrong link produces a wrong action, not just a scoring error).

\paragraph{Long-horizon agent memory and state management.}
Generative Agents~\cite{park2023generative} maintain consistent behavior via observation, reflection, and retrieval. Voyager~\cite{wang2023voyager} persists a skill library across episodes. \citet{sumers2024cognitive} formalize cognitive architectures for language agents, identifying memory as a core module but noting most agents lack mechanisms for detecting when stored state has become stale. AutoGen~\cite{wu2023autogen} passes state between agents via message history with no mechanism for detecting entity drift. Our work identifies a specific memory failure: the referent slot can silently bind to a different entity even when the agent ``remembers'' it resolved something earlier. The failure is not in retrieval (the agent recalls the prior binding) but in validity (the recalled binding is wrong or stale).

\paragraph{Belief-state repair and error recovery.}
Reflexion~\cite{shinn2023reflexion} uses verbal self-reflection to improve planning across episodes. Self-Refine~\cite{madaan2023selfrefine} iterates on outputs using self-feedback. CRITIC~\cite{gou2024critic} enables tool-grounded verification, showing that external checking outperforms self-correction. Our oracle re-verification is conceptually aligned with CRITIC's tool-grounded checking: it re-derives the binding from the original description using an external resolver. The key distinction: the error to be recovered is not in \emph{reasoning} (the agent's logic may be sound) but in \emph{reference} (the entity has silently changed). Standard self-reflection cannot detect this because internal state appears consistent.

\paragraph{Fault injection methodology.}
Our injection protocol draws on a long tradition of fault injection in dependability engineering~\cite{hsueh1997faultinjection,natella2016}. Chaos engineering~\cite{basiri2016chaos} applies controlled failure injection to production systems to verify resilience. We adapt this to LLM-agent evaluation: rather than injecting hardware faults, we inject a known-wrong entity binding and observe whether defenses propagate or correct it. This gives a capability-independent sample for measuring propagation behavior, decoupled from each model's natural error rate.

\paragraph{Multi-step and trajectory-level agent evaluation.}
AgentBench~\cite{liu2023agentbench} evaluates LLMs across eight environments with binary task success. WebArena~\cite{zhou2024webarena} measures multi-step web tasks where entity tracking is implicitly critical. MINT~\cite{wang2024mint} evaluates multi-turn tool-use with language feedback. $\tau$-bench~\cite{yao2024taubench} introduces pass$^k$ reliability. Trajectory-level diagnostics~\cite{traject2025} decompose task success into per-step decisions. All score whether a task \emph{succeeded}, not whether the agent maintained the same entity reference across steps. Our metrics ask a more specific question: across steps that are individually reasonable (correct tool, plausible entity), did the agent maintain referential identity?

\paragraph{Tool-use safety and high-risk evaluation.}
ToolEmu~\cite{ruan2024toolemu} emulates risky tool-use environments, categorizing harms including privacy breaches from wrong-person targeting. These harm categories align with our wrong-entity failure mode, but ToolEmu evaluates single tool calls rather than measuring how errors compound across a workflow. Our injection results quantify the compounding: a single wrong binding produces an average of 2.6 wrong actions per affected workflow under the entity lock.

\paragraph{Long-context state degradation.}
\citet{liu2024lostmiddle} showed that LLMs access information best at the beginning or end of input, with substantial degradation in the middle. \citet{shi2023irrelevant} demonstrated that irrelevant context degrades reasoning, analogous to our distractor-injection pattern. These findings provide a mechanistic explanation for binding drift: as the workflow progresses, the original resolution is positionally disadvantaged and surrounded by distractors. Our per-step error curve (8\% rising to 28\%) is consistent with this positional-degradation hypothesis, though our 3--8 step workflows are far shorter than the contexts studied in that literature.

\section{Limitations}
\label{sec:limitations}

\textbf{Ecological validity.} Workflows are 3--8 steps on synthetic entity stores. Real enterprise workflows (15--50 steps, noisy records, concurrent state changes, multiple providers) may exhibit different drift dynamics. Reported rates characterize failure behavior under controlled pressure, not deployment prevalence. All models are accessed through a single provider (Amazon Bedrock); other providers may behave differently.

\textbf{Binding-state coverage.} The benchmark exercises at most one derived slot per workflow (D3). Richer cascades (e.g., account $\to$ owner $\to$ owner's calendar $\to$ conflicting event) remain untested.

\textbf{Injection protocol.} The lock-amplification result under injection is partly mechanistic: the empirical contribution is the magnitude ($3.0\times$ aggregate, up to $8.5\times$) and the per-model variation, not the qualitative direction. The benchmark annotates three injection-phrasing variants but the current experiment uses a single fixed phrasing; injection position is fixed at step~1; and more gradual corruption patterns (stale records, conflicting metadata) are not modeled. The system prompt instructs agents not to ``silently switch,'' which may inflate lock compliance; a different prompt could shift absolute rates.

\textbf{Construct and statistical scope.} Our drift definition does not distinguish silent re-resolution from context-length forgetting (both score identically). Labeling clarification as ``over-clarification'' under injected state-conflict may be too harsh (it may be rational). Bootstrap 95\% CIs are reported but no formal hypothesis tests. The $3.0\times$ aggregate may obscure model-specific effects (variation $1.8$--$8.5\times$). Workflows within a pattern share structural templates, reducing effective independence.

\section{Conclusion}
\label{sec:conclusion}

Persistence and re-verification are not the same thing. An entity lock eliminates drift but, under controlled injection, amplifies propagation up to $8.5\times$ on the most affected model (Claude Opus~4.5), with the effect holding on all 8 backends tested ($3.0\times$ aggregate). A practical LLM cross-verifier (a single cheap second call using a small model) reduces propagation by 79\%, closing the gap to within 1 percentage point of the oracle upper-bound (80\%). For system builders: ``resolve once and carry forward'' amplifies errors when the first resolution is wrong. A cheap independent re-verification call before high-risk actions is sufficient to break the cascade.

\paragraph{Acknowledgments.}
This work was conducted in the authors' personal capacity and did not receive external funding. The benchmark, defenses, harness, and per-run trajectory logs are available at \url{https://github.com/shashank-indukuri/binding-drift}. All data is synthetic.

\bibliographystyle{unsrtnat}
\bibliography{references}

\appendix

\section{Representative Workflow Examples}
\label{app:examples}

We show two representative workflows verbatim to illustrate multi-step structure and how drift/propagation manifest.

\paragraph{D1 Temporal recurrence (calendar, 5 steps).}
\begin{quote}
\textbf{Entities:} 4 events all named ``Weekly Sync'' (customer-facing, internal-eng, executive, all-hands). Target: \texttt{evt\_sync\_customer}.\\
\textbf{Step 1:} ``Find the customer-facing Weekly Sync.'' $\to$ search (resolve)\\
\textbf{Step 2:} ``Add a note to it that the agenda is locked.'' $\to$ search\\
\textbf{Step 3:} ``Confirm its attendee list.'' $\to$ search\\
\textbf{Step 4:} ``Reschedule that meeting to 4pm.'' $\to$ \texttt{reschedule\_event} (gold: \texttt{evt\_sync\_customer})\\
\textbf{Step 5:} ``Then cancel it and mark it moved.'' $\to$ \texttt{cancel\_event} (gold: \texttt{evt\_sync\_customer})\\
\textbf{Drift risk:} Steps 4--5 say ``that meeting'' / ``it.'' The disambiguating qualifier ``customer-facing'' is never repeated.
\end{quote}

\paragraph{D3 Dependent chain (CRM $\to$ email, 4 steps).}
\begin{quote}
\textbf{Entities:} 2 accounts (Northwind Global, Northwind Labs) + 2 people (Dana=owner of Global, Sam=owner of Labs). Target: \texttt{acct\_northwind\_global}.\\
\textbf{Step 1:} ``Pull up the Northwind Global enterprise account.'' $\to$ search (resolve)\\
\textbf{Step 2:} ``Add a call note to it: renewal in progress.'' $\to$ \texttt{add\_call\_note} (gold: \texttt{acct\_northwind\_global})\\
\textbf{Step 3:} ``Update its renewal date to next quarter.'' $\to$ \texttt{update\_customer\_record} (gold: \texttt{acct\_northwind\_global})\\
\textbf{Step 4:} ``Email that account's owner a summary.'' $\to$ \texttt{send\_email} (gold: \texttt{person\_dana})\\
\textbf{Propagation risk:} If step~1 mis-resolves to Northwind Labs, step~4 emails Sam (Labs' owner) instead of Dana. This is a cross-domain cascade.
\end{quote}

\paragraph{Trajectory visualization: drift vs.\ propagation on D3.}
Table~\ref{tab:trajectory} traces a single D3 workflow under three conditions to show how the same workflow produces different failure modes depending on the initial binding and the defense.

\begin{table}[h]
\caption{Step-by-step trace of a D3 (dependent-chain) workflow under three conditions. Target: \texttt{acct\_northwind\_global}. The lock prevents drift (middle column) but amplifies propagation when the initial binding is wrong (right column).}
\label{tab:trajectory}
\centering
\small
\begin{tabular}{lp{3.2cm}p{3.2cm}p{3.5cm}}
\toprule
Step & Baseline (no defense) & Lock (correct start) & Lock + Injection (wrong start) \\
\midrule
1 (resolve) & Binds \texttt{global} \cmark & Binds \texttt{global}, lock set \cmark & Lock pre-set to \texttt{labs} \xmark \\
2 (add note) & Acts on \texttt{global} \cmark & Lock enforces \texttt{global} \cmark & Lock enforces \texttt{labs} \xmark \\
3 (update) & Acts on \texttt{global} \cmark & Lock enforces \texttt{global} \cmark & Lock enforces \texttt{labs} \xmark \\
4 (email owner) & Emails Dana \cmark & Emails Dana \cmark & Emails Sam (Labs' owner) \xmark \\
\midrule
Result & \textbf{Success} & \textbf{Success + no drift} & \textbf{3 wrong scored actions (cascade)} \\
\bottomrule
\end{tabular}
\end{table}

\section{Per-Pattern Analysis}
\label{app:patterns}

Table~\ref{tab:by_pattern} reports wrong-entity actions per drift pattern under injection (baseline and lock configs, 8 models aggregated). The dependent-chain (D3) pattern shows the strongest lock effect ($16.6\times$): baseline produces 61 wrong actions, but under the lock the seeded wrong binding cascades into 1{,}011 wrong actions because a wrong account binding propagates into emailing the wrong person at the final step. The distractor-injection pattern (D2) shows $2.9\times$ amplification. Temporal (D1) and silent-switch (D5) show $1.5{-}2.0\times$. True-ambiguity (D4) is neutral with respect to the lock ($1.0\times$), not because models consistently clarify, but because both baseline and lock configurations frequently guess despite the unresolvable target (571 and 565 wrong actions respectively).

\begin{table}[h]
\caption{Wrong-entity actions by drift pattern under injection (baseline vs.\ lock, 8 models aggregated, N=320 workflows per pattern-config cell).}
\label{tab:by_pattern}
\centering
\begin{tabular}{lrrr}
\toprule
Pattern & Baseline & Lock & Lock/$\,$Baseline \\
\midrule
D1 Temporal & 194 & 383 & $2.0\times$ \\
D2 Distractor injection & 245 & 722 & $2.9\times$ \\
D3 Dependent chain & 61 & 1{,}011 & $16.6\times$ \\
D4 True ambiguity$^\dagger$ & 571 & 565 & $1.0\times$ (clarifies) \\
D5 Silent switch & 407 & 630 & $1.5\times$ \\
\bottomrule
\multicolumn{4}{l}{\footnotesize $^\dagger$True-ambiguity: gold=CLARIFY; any ACT counts as wrong. High baseline reflects}\\
\multicolumn{4}{l}{\footnotesize models guessing despite ambiguity, not the lock mechanism.}\\
\end{tabular}
\end{table}

\section{Compute Environment and Model Details}
\label{app:compute}

All models are accessed as hosted endpoints through the Amazon Bedrock runtime (Converse API) in the \texttt{us-east-1} region; no model weights are run locally. Table~\ref{tab:models} lists the exact Bedrock model identifiers used.

\begin{table}[h]
\caption{Model backends and Bedrock identifiers. Temperature is not explicitly set; Bedrock applies model-specific defaults (typically greedy or near-greedy for instruction-tuned models). Setting temperature=0 reduces sampling variability but does not guarantee byte-identical outputs across provider-side revisions.}
\label{tab:models}
\centering
\small
\begin{tabular}{lll}
\toprule
Model (paper name) & Bedrock \texttt{modelId} & Family \\
\midrule
Nova Micro & \texttt{us.amazon.nova-micro-v1:0} & Amazon \\
Llama-3.1-8B & \texttt{us.meta.llama3-1-8b-instruct-v1:0} & Meta \\
Nova 2 Lite & \texttt{us.amazon.nova-2-lite-v1:0} & Amazon \\
Claude Haiku 4.5 & \texttt{us.anthropic.claude-haiku-4-5-20251001-v1:0} & Anthropic \\
Llama-3.3-70B & \texttt{us.meta.llama3-3-70b-instruct-v1:0} & Meta \\
Nova Premier & \texttt{us.amazon.nova-premier-v1:0} & Amazon \\
Claude Sonnet 4.5 & \texttt{us.anthropic.claude-sonnet-4-5-20250929-v1:0} & Anthropic \\
Claude Opus 4.5 & \texttt{us.anthropic.claude-opus-4-5-20251101-v1:0} & Anthropic \\
\bottomrule
\end{tabular}
\end{table}

Each call uses \texttt{maxTokens}=700, sequential execution with 0.15--0.8s inter-request delay (larger models use longer delays to avoid Bedrock connection-limit throttling). Experiments were run July 6--10, 2026.

The experiment harness, task generator, and scoring are written in Python~3.9 and depend on \texttt{boto3} (1.42.97), \texttt{botocore} (1.42.97), \texttt{pandas} (2.3.3), \texttt{tenacity} (9.1.2), and \texttt{tqdm} (4.68.3). The harness is deterministic given model outputs; the task suite and scoring use no random number generation.

\section{Prompt Format}
\label{app:prompt}

Each model call uses the Bedrock Converse API with a system message and one user message. The system message is fixed across all steps:

\begin{quote}
\small
\texttt{You are a tool-augmented enterprise agent operating over MULTIPLE STEPS. Across the whole task you act on the SAME real-world entities. When a later step refers to ``it'', ``that'', ``the account'', etc., it means the entity you already identified earlier. Do not silently switch to a different entity.}

\texttt{Output ONLY valid JSON per step: \{``decision'': ``ACT'' | ``CLARIFY'' | ``DEFER'', ``tool'': ``tool\_name\_or\_null'', ``bindings'': \{``slot'': ``entity\_id''\}, ``clarification\_question'': ``string\_or\_null'', ``provenance'': [...]\}}
\end{quote}

The user message for each step is a JSON payload containing: the defense-config hint, workflow ID, domain, step instruction, available tools (with \texttt{requires} annotations), candidate entities (with metadata), required binding slots, and the prior-step history (instruction + decision + bindings for each completed step). Under entity-lock configs, the payload additionally includes \texttt{``locked\_entities'': \{slot: id\}}. Under provenance configs, recent provenance records are included.

\paragraph{Injection format.} Under controlled injection, a synthetic prior turn is prepended to the history:
\begin{quote}
\small
\texttt{\{``instruction'': ``(prior) identified the [slot] as [wrong\_id]'', ``decision'': ``ACT'', ``bindings'': \{``[slot]'': ``[wrong\_id]''\}\}}
\end{quote}
where \texttt{[slot]} is the workflow's carry slot (e.g., ``account'') and \texttt{[wrong\_id]} is the annotated distractor entity ID (e.g., ``acct\_northwind\_labs''). This is placed as the most recent history entry before step~1's instruction, simulating a prior action that resolved the wrong entity. The benchmark annotates three injection-phrasing variants (authoritative, uncertain, tool-return) for future sensitivity analysis, but the current experiment uses a single fixed phrasing for all workflows; stronger or weaker formulations may modulate propagation rate.

\paragraph{LLM re-verification protocol.} Under the \texttt{llm\_self\_reverify} and \texttt{llm\_cross\_reverify} configs, a second Bedrock Converse call is made at each action step after the primary agent call. The re-verifier receives only the \emph{original} step-1 referent phrase and the full candidate-entity list:

\begin{quote}
\small
\texttt{You are an independent entity resolver. Given the original user request and candidate entities, determine which entity the request refers to.}

\texttt{Original request: ``\{step1\_referent\}''}

\texttt{Candidate entities: [... entity list with metadata ...]}

\texttt{Which entity ID does the original request refer to? Return ONLY: \{``entity\_id'': ``the\_correct\_id''\}. If genuinely ambiguous, return: \{``entity\_id'': ``AMBIGUOUS''\}.}
\end{quote}

If the re-verifier's resolved ID disagrees with the agent's carried binding, the binding is overwritten with the re-verifier's answer and the lock is updated. If the re-verifier returns ``AMBIGUOUS'' or fails to parse, no correction is applied. Under \texttt{llm\_self\_reverify}, the same model that acts as the agent also serves as the re-verifier. Under \texttt{llm\_cross\_reverify}, Nova Micro (\texttt{us.amazon.nova-micro-v1:0}) serves as the verifier regardless of which model is the agent. The cross-verifier adds one cheap API call per action step (${\sim}$0.3s latency, minimal token cost).

\section{Offline Validation}
\label{app:validation}

Before spending compute on live models, we validated the scoring harness using four scripted fake agents with deterministic known behaviors:

\begin{itemize}[nosep]
\item \textbf{Oracle:} always binds the ground-truth entity; clarifies on true-ambiguity workflows.
\item \textbf{Drifter:} correct at step~1; switches to a distractor at the \emph{last} action step only.
\item \textbf{Propagator:} binds a distractor at step~1 and carries it forward every step.
\item \textbf{Guesser:} on true-ambiguity workflows, ACTs (guesses) instead of clarifying.
\end{itemize}

25 assertions verify the following categories:
\begin{itemize}[nosep]
\item \textbf{Drift scoring (5):} oracle has zero drift; drifter triggers drift=1 with exactly 1 wrong action; drift is not counted as propagation.
\item \textbf{Lock behavior (4):} lock pins the first binding; lock prevents drifter's late switch; lock under injection propagates the wrong entity.
\item \textbf{Gate behavior (3):} gate fires CLARIFY on divergence from lock; no wrong action passes the gate.
\item \textbf{Propagation scoring (5):} propagator is mis-bound at step~1; propagated flag set; compounding factor equals number of action steps; no recovery.
\item \textbf{Oracle re-verify (4):} under injection, reverify corrects the wrong binding to zero wrong actions; reverify does not break the clean oracle case; reverify does not falsely correct true-ambiguity.
\item \textbf{True-ambiguity control (2):} guesser fails; oracle correctly clarifies.
\item \textbf{Disjoint scoring (2):} a workflow cannot contribute to both drift and propagation metrics simultaneously.
\end{itemize}

All 25 assertions pass. The test file is released as \texttt{code/test\_offline.py}.

\section{Code and Data Release}
\label{app:release}

All code, data, and results are publicly available at:
\begin{center}
\url{https://github.com/shashank-indukuri/binding-drift}
\end{center}

The repository contains:
\begin{itemize}[nosep]
\item \texttt{code/generate\_hard\_workflows.py}: workflow generator (produces \texttt{data/workflows\_hard.jsonl})
\item \texttt{code/environment.py}: stateful environment, defense layers, per-step and trajectory scoring
\item \texttt{code/run\_drift\_experiment.py}: Bedrock multi-step runner (\texttt{-{}-inject} flag for injection mode)
\item \texttt{code/aggregate\_results.py}: trajectory metrics to summary CSVs and LaTeX tables
\item \texttt{code/aggregate\_injection.py}: injection-specific aggregation and capability ladder
\item \texttt{code/test\_offline.py}: 25-assertion offline validation (no API required)
\item \texttt{data/workflows\_200.jsonl}: all 200 workflows (deterministic, regenerable)
\item \texttt{results/}: per-step and per-workflow CSVs for all 40,000+ model calls (injection mode, 8 models, 5 configs, 200 workflows)
\end{itemize}

\paragraph{Reproduction modes.}
\begin{enumerate}[nosep]
\item \textbf{Tables from saved results (no API):} \texttt{python aggregate\_injection.py --wf results/workflows\_inject\_ALL.csv --outdir out/}
\item \textbf{Offline metric validation (no API):} \texttt{python test\_offline.py} (25/25 must pass)
\item \textbf{Full re-run (needs Bedrock access):} \texttt{python run\_drift\_experiment.py --workflows data/workflows\_hard.jsonl --out results/steps.csv --wf-out results/workflows.csv --models [IDs] --configs baseline lock ... reverify [--inject]}
\end{enumerate}

\end{document}